# Socially Interactive Agents for Preserving and Transferring Tacit Knowledge in Organizations

Requirements and approaches to AI-supported knowledge transfer

Martin Benderoth[1], Patrick Gebhard[2], Christian Keller[3], C. Benjamin Nakhosteen[4], Stefan Schaffer[2], Tanja Schneeberger[2]

[1] ad-artists GmbH, Kassel, Germany
[2] German Research Center for Artificial Intelligence (DFKI), Germany
[3] synartIQ GmbH, Bielefeld, Germany
[4] thyssenkrupp Steel Europe AG, Duisburg, Germany

## 1 Introduction

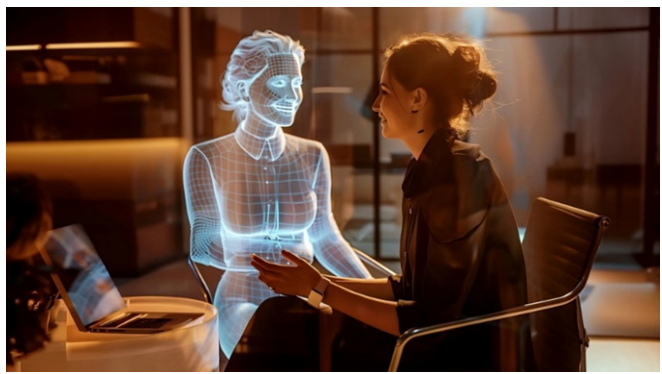

Figure 1: Vision of an AI preservation

In times of demographic change, a shortage of professionals and the increasing importance of knowledge as a key value creation factor, it is essential for companies to secure and pass on the valuable experiential knowledge of their employees (Tachkov & Mertens, 2016). There is a broad consensus that organizations must take measures to promote the transfer of their employees' knowledge. However, what is less often reflected upon is how difficult it is to achieve this in practice and that the preservation of knowledge gained from experience is a supreme discipline that goes far beyond simply taking notes of conversations. Companies that systematically pursue knowledge transfer generally leave the detailed tasks of recording and documenting hidden knowledge to the employees directly involved in the process and sometimes to a so-called knowledge transfer facilitator, who acts as a coach with methodological expertise to mediate between experts sharing and learners acquiring knowledge (Hofer-Alfeis, 2008).

Though valuable, this resource-intensive process lacks scalability, highlighting the need for innovative solutions that balance efficiency and effectiveness. Therefore, this paper presents a new direction to address the significant challenges in preserving and transferring tacit knowledge: Socially Interactive Agents (SIAs) (Lugrin, 2021) in the role of knowledge transfer facilitators.

In the following, we will present the scientific background on tacit knowledge management and transfer, as well as on SIAs. We will argue why SIAs are suited as knowledge transfer facilitators and how this could be realized technically. Afterwards, we present possible application scenarios and discuss challenges.

## 2 Background

### 2.1 Tacit Knowledge Management and Transfer

Knowledge Management is *"the process of applying a systematic approach to the capture, structure, management, and dissemination of knowledge throughout an organization in order to work faster, reuse best practices, and reduce costly rework from project to project"* (Nonaka et al., 2001).

Compared to explicit knowledge, which can be codified into something that is formal, structured and systematic, tacit knowledge, is highly personal and difficult to formalize. It is based on actions and experiences of an individual collected in different situations in specific practical contexts (Joia & Lemos, 2010). Tacit knowledge accounts for a significant proportion of knowledge in organizations (Hurley & Green, 2005; Nonaka & Takeuchi, 1995, Polanyi, 1966; Wah, 1999). Tapping into this knowledge holds enormous potential for organizations. Tacit knowledge is a special category of knowledge that is subjective, experience-based, and difficult to articulate. It is deeply embedded in the actions, experiences, ideals, values, and feelings of employees. While manual labor is often cited as an example of where tacit knowledge is used, employee performance in the much-quoted knowledge society also relies on a significant amount of tacit knowledge, e.g.:

– Expertise in complex decision-making processes: Specialists and managers use their tacit knowledge to make intuitive but well-founded decisions, especially when the available information is incomplete.
– Creativity and innovation: The unconscious connections that characterize tacit knowledge play an important role in the development of new ideas and solutions.
– Communication: Understanding nonverbal signals and being able to read between the lines to collaborate effectively is based on tacit knowledge.
– Problem-solving in unforeseen situations: Experienced employees draw on their tacit knowledge to react quickly and flexibly to new challenges and to develop solutions.

Traditional methods of knowledge retention, e.g., written documentation and passing on knowledge in training courses, are often insufficient for transferring such skills. These methods primarily emphasize explicit knowledge (e.g., figures, data, facts) and do not facilitate access to the deeper, often unconscious knowledge of experts. Here, a lack of knowledge about the possible handling of experiential knowledge results in an implementation deficit in knowledge retention. Previous approaches to knowledge transfer have attempted to bring people together so that they can access hidden knowledge in a joint dialog with trained knowledge transfer facilitators. In addition, tandems of experienced knowledge holders and newcomers are formed so that successors become familiar with the activities of the experts and gradually build up their own expertise and ability to act. Traditional instruments of personnel development and knowledge management such as job rotation, mentoring, lessons learned, storytelling, and communities of practice are also used. The methods applied to date are person-centered in the sense of individual support and, therefore, involve a great deal of effort. To adequately address the high professional and social complexity of knowledge transfer in the future and to cope with the enormous volume of knowledge retention required, new approaches need to be developed. These approaches should be designed to be resource-efficient and scalable to suit the size of the organization in question.

**2.2 Socially Interactive Agents as Knowledge Transfer Facilitators**

Socially Interactive Agents (SIAs) are virtually or physically embodied agents that are capable of autonomously communicating with people and each other in a socially intelligent manner using multimodal behaviors (verbal, paraverbal, and nonverbal) known from human-human interaction (Lugrin, 2021, 1). The initial goal of applying SIAs in the field of human-computer interaction was to enable human users to interact via communication channels that come naturally to them (Cassell et al., 2000). The transfer from communication styles that are known from human face-to-face interaction to the interaction with machines results in a more human-like interface that is intuitive to understand and to interact with. Beyond usability, scholars have shown that adding a human-like SIA to the computer interface results in a more pronounced social behavior towards computers (Krämer et al., 2018). Numerous studies yield social effects, demonstrating that humans' reactions towards virtual agents are remarkably similar to those towards human interlocutors.

SIAs evoke communication behavior in humans that is equivalent to that expected in a face-to-face conversation. This includes human-like communication strategies, cooperative behavior (Kopp et al., 2005), polite behavior (Hoffmann et al., 2009), and rapport building (Gratch et al., 2007). Beyond human-like conversational behavior SIAs induce affective responses that are known from human-human interaction (Schneeberger et al., 2019a, Schneeberger et al., 2019b, Schneeberger et al., 2023). The presence of a SIA leads to increased trust in an AI-system (Weitz et al., 2021). One reason for the high trust of humans in SIAs might be that they are experienced as supportive and safe interaction partners. Especially for topics where people have challenges to self-disclose in front of another human, several studies showed that SIAs evoked lower fear of self-disclosure (Bickmore et al. 2020; Lucas et al., 2014; Lucas et al., 2017). Krämer et al. (2018) found evidence that interactions with a virtual agent are experienced as socially rewarding and can meet social needs like human-human interactions. Pauw et al. (2022) showed that after talking about two negative emotions, anger and worry, and getting emotional and cognitive support, participants felt better—the target emotion was reduced, and the affect was generally improved. This led the authors to the conclusion that talking to a SIA can be socio-emotionally supporting and a valuable form of support at times of distress.

Several applications of SIAs in the fields of digital health (Bickmore, 2022) and digital business make use of human social behavior towards them. The latter category involves SIAs simulating experts, e.g., in job interview training (Gebhard et al., 2018), to support flow experiences and well-being at work (Beyrodt, 2023), or for reintegration (Gebhard et al., 2019).

To date, there has been no research examining the application of SIAs as knowledge transfer facilitators. However, based on the existing findings, SIAs have the potential to engage employees in empathic, natural-language dialogues, enabling them to externalize insights that might otherwise remain unspoken. The efficacy of these measures is contingent upon the establishment of trust, as employees often exhibit reluctance to share tacit knowledge without assurance that their contributions will be valorized and safeguarded.

## 3 Solution Approach

In the following, a new approach (Fig. 2) to tapping into tacit knowledge will be discussed. Although technology is used in the proposed solution, the focus is on people possessing experience as individuals with attitudes, opinions, and beliefs. In addition, there are SIAs, as a colleague that accompanies employees and helps them to articulate and share their expertise, experiences, and insights. So, instead of limiting themselves to traditional, sometimes unmotivating sources of information such as instructions, databases, or static training events, such empathic mentors could support the transfer of knowledge in day-to-day operations.

Building trust is of paramount importance for this approach, especially with regard to tacit knowledge. People are often reluctant to share such knowledge. In view of this, it is essential that SIAs for knowledge transfer are designed and developed accordingly. A SIA must be able to adapt to the individual needs of their mentee, provide feedback, and convey recognition, enabling a bidirectional relationship in which employees feel valued for their experience and knowledge. According to psychological research, warmth and competence play a significant role in building trust (Fiske, Cuddy, & Glick, 2007). It is, therefore, not a question of simply responding to the challenges of knowledge transfer with a maximum use of technology. Rather, socio-technical systems should be set up in which people act and interact with technology in protected spaces and in an intuitive manner.

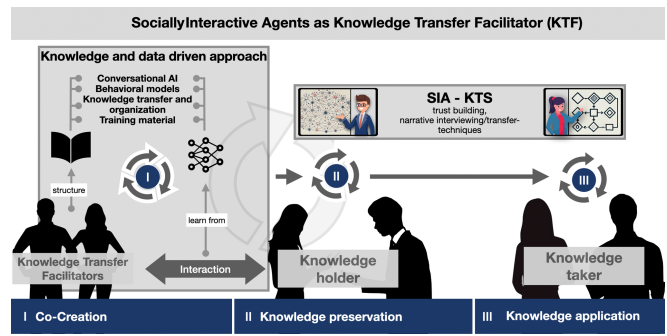

Figure 2: Approach for introducing SIAs as knowledge transfer facilitators

Key technology components include Large Language Models (LLMs) to generate contextually relevant dialog, Retrieval-Augmented Generation (RAG) to incorporate organizational knowledge into conversations, and Chain-of-Thought (CoT) prompts to guide users through structured reflection. These technologies enable SIAs to function not just as question-answering tools, but as active participants in knowledge elicitation, probing for implicit assumptions and connecting insights to broader organizational contexts.

The realization requires both 1) the combination of automatic speech recognition and speech generation with LLMs to record, process, and generate natural language as well as 2) technologies that can acquire domain-related context knowledge externally (retrieval augmented generation, RAG) and incorporate it into dialogs with users. The task of LLMs is to provide intuitively usable dialog interfaces that create a setting for users that feels similar to a conversation with colleagues at eye level (Gürcan, 2024). The SIA should be perceived as an interaction partner who can receive information on a wide variety of channels in real-time (e.g., also visually), is familiar with the context of interest, and has sufficient expertise to conduct a sophisticated dialog that corresponds to the user's level in the relevant field.

However, it is not enough for the LLMs to respond to user prompts in the way that is the case in today's standard, publicly available systems. In order to access the critical tacit knowledge of the users, they must also follow a dialog strategy, which can be implemented using chain-of-thought prompting (CoT). The dialog must be steered in a direction that allows users to unfold their deep experiential knowledge in conversation. SIAs could use CoT prompting to ask the right questions in the correct order, just as an experienced interviewer would (Wei et al., 2022). This can reveal knowledge and externalize thinking processes that users would otherwise hardly be aware of. Finally, with the help of RAG, SIAs can access company databases and other sources of information in real-time to link employees' experiences with relevant documents, company guidelines, and documented experiences of other employees (Gao et al., 2023). This approach has the potential to close the existing gaps between today's knowledge management systems and the still largely undiscovered field of AI-driven knowledge capture from a knowledge transfer perspective.

## 4 Application scenarios

The potential fields of application for the AI-based socially interactive knowledge transfer facilitators presented here are manifold. One example is the onboarding phase of new employees. The personalized SIA could guide newly hired employees through the company's process landscape and prepare their introduction to key personnel. Instead of being overwhelmed by the size and complexity of a new organization, employees would have access to an AI that is available at all times to guide them, answer questions, identify development steps, and suggest relevant training courses. The approach described could significantly shorten the period of onboarding and enable new employees to become well-integrated and productive team members more quickly.

Another use case is the knowledge retention of retiring experienced employees mentioned at the beginning. A SIA is able to enter into a dialog with such senior-level employees and ask them in-depth questions. These are not only aimed at determining what knowledge the experienced employees have, but also how they apply their knowledge in practice. The focus is, therefore, not exclusively on facts. Instead, it is about know-how, experience, judgements, and decisions. This opens up new opportunities for a digital legacy for individuals and organizations to address the challenges of potential knowledge loss.

A SIA can also be used as part of career-biographical personnel development. A personal SIA could document the employee's development progress, identify areas for development and recommend personalized training and development measures. In this case, the SIA acts as a personal AI coach who supports individual career development.

These application examples show that employees' trust in their SIA is a decisive success factor in all conceivable scenarios. If employees do not trust the solution, even the most advanced and best AI will remain unused. In this case, it can be assumed that employees will also be unwilling to share their valuable and highly personal experiences with the AI tool. In order to generate the necessary trust, a multi-level approach is proposed. One core element is transparency. Employees must be informed about how SIAs work, the data they collect, and how this data is used. The second element is personalization. SIAs are adapted to the individual needs and preferences of users. This circumstance creates a motivating user experience that users find pleasant. Furthermore, feedback functions are implemented that allow users to provide input on how their SIA should develop over time. SIA and employees thus interact in a bidirectional relationship in which both learn from each other in continuous feedback loops.

## 5 Critical discussion

The risks of both the general, company-wide, and project-specific development of AI solutions should not be concealed in this context but addressed openly. It must be determined whether there is a risk of job losses due to the continuous increase in efficiency. The misuse of personal data must be prevented in all developed solutions, particularly in the case of permanently attentive and observing AI agents. In addition, it must be investigated whether new forms of unequal treatment or bias can arise, for example, through the incorrect training of AI models that is not taken into account. Generative AI is based on probabilities and patterns in training data. It delivers the statistically most probable but not necessarily the correct or best solution. This is particularly evident in complex decision-making processes that require a critical examination of the context. It is, therefore, necessary to incorporate relevant information through reasoning in order to scrutinize the AI's output, reject it if necessary, and revise it.

These valid concerns must be taken into account in projects to implement SIAs by developing and implementing solutions responsibly. In addition to compliance with IT security standards, this implies, in particular, the consideration of ethical, psychological, and sociological framework conditions both in the project design and in the composition of consortia. The overarching goal is to develop fair solutions that consider not only technical feasibility but also the broader effects and ethical implications for the workforce and society. Early clarification of such issues is of fundamental importance before generative technologies are so deeply anchored in operational reality that it is no longer possible to correct course.

## 6 Conclusion

In this paper, we proposed the application of SIAs as knowledge transfer facilitators in organizations to preserve and transfer tacit knowledge for the rapidly growing knowledge management market. There are no known systems that integrate AI-based SIAs and immersive technologies into solutions that build on such a deep understanding of the psychological relationships in knowledge transfer. Although the development of knowledge transfer-related SIAs is still at an early stage of research, the enormous potential for a disruptive change in professional knowledge management is already becoming apparent.

## References

Beyrodt, S., Nicora, M. L., Nunnari, F., Chehayeb, L., Prajod, P., Schneeberger, T., André, E., Malosio, M., Gebhard, P. & Tsovaltzi, D. (2023). Socially interactive agents as cobot avatars: Developing a model to support flow experiences and well-being in the workplace. In *Proceedings of the 23rd ACM International Conference on Intelligent Virtual Agents* (pp. 1–8). https://doi.org/10.1145/3570945.3607349

Bickmore, T. W. (2022). *Health-related applications of socially interactive agents*. In B. Lugrin, C. Pelachaud, & D. Traum (Eds.), The handbook on socially interactive agents: 20 years of research on embodied conversational agents, intelligent virtual agents, and social robotics. Volume 2: Interactivity, platforms, application (pp. 403–436). Association for Computing Machinery. https://doi.org/10.1145/3563659.3563672

Bickmore, T. W., Rubin, A., & Simon, S. R. (2020). Substance use screening using virtual agents: Towards automated screening, brief intervention, and referral to treatment (SBIRT). *Proceedings of the 20th International Conference on Intelligent Virtual Agents*, 1–7. https://doi.org/10.1145/3383652.3423869

Cassell, J., Sullivan, J., Churchill, E., & Prevost, S. (2000). *Embodied conversational agents*. MIT Press.

Fiske, S. T., Cuddy, A. J., & Glick, P. (2007). Universal dimensions of social cognition: Warmth and competence. *Trends in cognitive sciences, 11*(2), 77–83.

Gao, Y., Xiong, Y, Gao, X., Jia, K., Pan, J., Bi, Y., Dai, Y., Sun, J., Wang, M., & Wang, H. (2023). *Retrieval-augmented generation for large language models*: A survey. arXiv preprint arXiv:2312.10997.

Gebhard, P., Schneeberger, T., André, E., Baur, T., Damian, I., Mehlmann, G., König, C. & Langer, M. (2018). Serious games for training social skills in job interviews. *IEEE Transactions on Games*, 11(4), 340–351. https://doi.org/10.1109/TG.2018.2808525

Gebhard, P., Schneeberger, T., Dietz, M., André, E. & Bajwa, N. u. H. (2019). Designing a mobile social and vocational reintegration assistant for burn-out outpatient treatment. In *Proceedings of the 19th ACM International Conference on Intelligent Virtual Agents* (pp. 13–15). https://doi.org/10.1145/3308532.3329460

Gratch, J., Wang, N., Gerten, J., Fast, E., & Duy, R. (2007). Creating rapport with virtual agents. *Proceedings of the 7th International Conference on Intelligent Virtual Agents*, 125–138. https://doi.org/10.1007/978-3-540-74997-4_12

Gürcan, Ö. (2024). LLM-Augmented Agent-Based Modeling for Social Simulations: Challenges and Opportunities. *HHAI 2024: Hybrid Human AI Systems for the Social Good*, (pp. 134–144).

Hofer-Alfeis, J. (2008). Knowledge management solutions for the leaving expert issue. *Journal of Knowledge Management, 12*(4), 44–54.

Hoffmann, L., Krämer, N. C., Lam-Chi, A., & Kopp, S. (2009). Media equation revisited: Do users show polite reactions towards an embodied agent? *Proceedings of the 9th International Conference on Intelligent Virtual Agents*, 159–165. https://doi.org/http://dx.doi.org/10.1007/978-3-642-04380-2_19

Hurley, T. A., & Green, C. W. (2005). Knowledge Management and the Nonprofit Industry. A Within and Between Approach. *Journal of Knowledge Management Practice*, January.

Joia, L. A., & Lemos, B. (2010). Relevant factors for tacit knowledge transfer within organisations. *Journal of Knowledge Management, 14*(3), 410–427.

Kopp, S., Gesellensetter, L., Krämer, N. C., & Wachsmuth, I. (2005). A conversational agent as museum guide: Design and evaluation of a real-world application. *Proceedings of the 5th International Conference on Intelligent Virtual Agents*, 329–343. https://doi.org/10.1007/11550617_28

Krämer, N. C., Lucas, G., Schmitt, L., & Gratch, J. (2018). Social snacking with a virtual agent—on the interrelation of need to belong and effects of social responsiveness when interacting with artificial entities. *International Journal of Human-Computer Studies, 109*(1), 112–121. https://doi.org/10.1016/j.ijhcs.2017.09.001

Lucas, G. M., Gratch, J., King, A., & Morency, L.-P. (2014). It's only a computer: Virtual humans increase willingness to disclose. *Computers in Human Behavior*, 37, 94–100. https://doi.org/10.1016/j.chb.2014.04.043

Lucas, G. M., Rizzo, A., Gratch, J., Scherer, S., Stratou, G., Boberg, J., & Morency, L.-P. (2017). Reporting mental health symptoms: Breaking down barriers to care with virtual human interviewers. *Frontiers in Robotics and AI*, 4, Article 51. https://doi.org/10.3389/frobt.2017.00051

Lugrin, B. (2021). Introduction to socially interactive agents. In B. Lugrin, C. Pelachaud & D. Traum (Hrsg.), *The handbook on socially interactive agents: 20 years of research on embodied conversational agents, intelligent virtual agents, and social robotics. Volume 1: Methods, behavior, cognition* (pp. 77–104). Association for Computing Machinery. https://doi.org/10.1145/3477322.3477326

Nonaka, I, & Takeuchi, H. (1995). *The Knowledge-Creating Company: How Japanese Companies Create the Dynamics of Innovation*. Oxford: Oxford University Press.

Nonaka, I., Toyama, R., & Konno, N. (2001). Seci, ba and leadership: a unified model of dynamic knowledge creation. In I. Nonaka, D. J. Teece (Hrsg.) *Managing industrial knowledge: creation, transfer and utilization* (pp. 14–43). SAGE Publications Ltd, https://doi.org/10.4135/9781446217573.n2

Pauw, L. S., Sauter, D. A., van Kleef, G. A., Lucas, G. M., Gratch, J., & Fischer, A. H. (2022). The avatar will see you now: Support from a virtual human provides socio-emotional benefits. *Computers in Human Behavior*, 136, Article 107368. https://doi.org/10.1016/j.chb.2022.107368

Polanyi, M. (1966). *The Tacit Dimension*. New York: Doubleday.

Schneeberger, T., Ehrhardt, S., Anglet, M. S., & Gebhard, P. (2019b). Would you follow my instructions if I was not human? Examining obedience towards virtual agents. In *Proceedings of the 8th International Conference on Affective Computing and Intelligent Interaction* (pp. 1–7). IEEE. https://doi.org/10.1109/ACII.2019.8925501

Schneeberger, T., Reinwarth, A. L., Wensky, R., Anglet, M. S., Gebhard, P., & Weßler, J. (2023). Fast Friends: Generating Interpersonal Closeness between Humans and Socially Interactive Agents. In *Proceedings of the 23rd ACM International Conference on Intelligent Virtual Agents* (pp. 1–8). https://doi.org/10.1145/3570945.3607302

Schneeberger, T., Sauerwein, N., Anglet, M. S. & Gebhard, P. (2021). Stress management training using biofeedback guided by social agents. In *Proceedings of the 26th International Conference on Intelligent User Interfaces* (pp. 564–574). https://doi.org/10.1145/3397481.3450683

Schneeberger, T., Scholtes, M., Hilpert, B., Langer, M., & Gebhard, P. (2019a). Can social agents elicit shame as humans do? In 2019 8th International Conference on Affective Computing and Intelligent Interaction (ACII) (pp. 164–170). IEEE. https://doi.org/10.1109/ACII.2019.8925481

Tachkov, P., & Mertens, M. (2016). Implizites Mitarbeiterwissen in KMU: Den Schatz stiller Wissensressourcen heben. *Wissensmanagement, 2*, 28–30.

Wah, L. (1999). Can knowledge be measured? *Management Review, 88*(5).

Wei, J., Wang, X., Schuurmans, D., Bosma, M., Ichter, B., Xia, F., Chi, E., Le, Q., & Zhou, D. (2022). Chain-of-thought prompting elicits reasoning in large language models. *Advances in neural information processing systems*, 35, 24824–24837.

Weitz, K., Schiller, D., Schlagowski, R., Huber, T., & André, E. (2021). "Let me explain!": Exploring the potential of virtual agents in explainable AI interaction design. *Journal on Multimodal User Interfaces*, 15 (2), 87–98. https://doi.org/10.1007/s12193-020-00332-0